\documentclass[conference]{IEEEtran}
\IEEEoverridecommandlockouts
\usepackage{cite}
\usepackage{amsmath,amssymb,amsfonts}
\usepackage{algorithmic}
\usepackage{graphicx}
\usepackage{textcomp}
\usepackage{xcolor}
\usepackage{booktabs}
\usepackage{url}
\def\BibTeX{{\rm B\kern-.05em{\sc i\kern-.025em b}\kern-.08em
    T\kern-.1667em\lower.7ex\hbox{E}\kern-.125emX}}

\begin{document}

\title{An Empirical Study of Entropy-Conserving Binarization in H.264/AVC CABAC}

\author{
\IEEEauthorblockN{Vinamra Singh}
\IEEEauthorblockA{\textit{Department of Informatics and Networked Systems} \\
\textit{University of Pittsburgh}\\
Pittsburgh, PA, USA \\
vis213@pitt.edu}
}

\maketitle

\begin{abstract}
CABAC, the entropy coder of H.264/AVC and the basis for the entropy coders of HEVC and VVC, decomposes multi-symbol values into binary decisions through a binarization scheme before passing them to a binary arithmetic coder. The H.264 standard uses Truncated Unary plus $k$-th order Exp-Golomb (UEG); alternatives include canonical Huffman codes and the entropy-conserving binarization scheme (ECB) of Srivastava, which provably preserves entropy when mapping $m$-ary data to $m-1$ binary strings but has not been empirically evaluated inside a production binary arithmetic coder. We integrate ECB into a from-scratch H.264/AVC CABAC implementation alongside UEG, canonical Huffman with a single shared context, and a Huffman variant with per-bin-position contexts (HuffmanPos), all sharing a common M-coder backend. We benchmark all four on synthetic distributions, quantized DCT residuals from a procedural test image, and the full 24-image Kodak True Color Image Suite ($2{,}480$ round-trip trials, bit-exact verified). On the procedural image we identify a sparsity-driven crossover at quantization step $Q=8$ where ECB overtakes single-context Huffman, reaching $27$ percentage points below at $Q=32$. On Kodak the crossover shifts below the tested range and ECB beats single-context Huffman at every $Q$, with the gap growing from $0.031$ to $0.113$ bits per symbol. HuffmanPos, which shares Huffman's codewords but allocates one M-coder context per bin position, beats ECB on $12$ of the $15$ source cells and loses by at most $0.56$ percentage points on the other three, despite committing the same number of bins per source symbol as single-context Huffman. This isolates the dominant mechanism: at low source entropy, the rate gap between schemes is driven primarily by context allocation strategy over the bin stream, not by the per-symbol bin count of the binarization. ECB's rate efficiency costs $7$ to $10\times$ in decoder latency on large alphabets, traced to an $O(N \cdot m)$ decoder implementation; we sketch an interleaved single-pass variant that would close this gap. Code, benchmarks, and raw data are open source.
\end{abstract}


\section{Introduction}
\label{sec:intro}

Modern video and image codecs depend on entropy coding to convert symbol streams into compact bit sequences. The H.264/AVC~\cite{wiegand2003h264} and HEVC~\cite{sullivan2012hevc} video standards adopted Context-Adaptive Binary Arithmetic Coding (CABAC)~\cite{marpe2003cabac} as their entropy coder, a design that carried forward through VVC~\cite{bross2021vvc} and remains the dominant approach in production video compression. CABAC operates on binary decisions: each multi-symbol value is first decomposed into a sequence of bins by a \emph{binarization scheme}, after which a binary arithmetic coder (the M-coder) compresses the bins using adaptive context models. The compressed rate is determined jointly by two factors: the bin distribution that the M-coder sees (a property of the binarization), and the conditioning structure available for context adaptation (a property of how contexts are allocated over the bin stream).

The H.264 standard binarization for residual coefficients combines a Truncated Unary (TU) prefix with a $k$-th order Exp-Golomb (EG-$k$) suffix, the UEG scheme. UEG was designed for the geometric-tailed distributions typical of DCT residuals and serves as the default across the standard. Alternative binarizations exist: canonical Huffman codes~\cite{huffman1952method} provide symbol-level optimality, and Srivastava~\cite{srivastava2014} introduced an \emph{entropy-conserving binarization} (ECB) that maps an $m$-ary source into $m-1$ binary strings while provably preserving the total entropy in linear time.

ECB has favorable theoretical properties but, to our knowledge, has not been empirically evaluated inside a production binary arithmetic coder. The entropy-conservation theorem guarantees lossless rate at the source-coding level; how the scheme behaves inside CABAC, where bin streams pass through adaptive context models, is a separate empirical question. A related question, which has received less attention in the empirical literature, is how much of the observed rate gap between binarizations is attributable to the binarization itself versus the context allocation strategy applied to the resulting bin stream.

This paper presents that evaluation. We integrate ECB into a from-scratch H.264/AVC CABAC implementation alongside UEG and two canonical Huffman variants: one with a single shared context across all bin positions (Huffman), and one with one context per bin position (HuffmanPos). The two Huffman variants share codewords but differ in context wiring, isolating the effect of context allocation from binarization choice. All four schemes route through a shared M-coder. We benchmark on synthetic distributions, quantized DCT residuals from a procedural test image, and the full 24-image Kodak True Color Image Suite. Our contributions are:

\begin{itemize}
    \item A reproducible CABAC implementation following ITU-T H.264 Section 9.3~\cite{itu2003h264}, with four pluggable binarizers (UEG, ECB, Huffman, HuffmanPos) sharing a single M-coder backend.
    \item A benchmark suite across synthetic sources, a 50-seed procedural DCT image sweep, and a sweep on all 24 Kodak natural photographs, with bit-exact round-trip verification on every trial ($2{,}480$ trials total).
    \item An empirical finding: a sparsity-driven crossover whose position depends on image content. On the procedural image the crossover between single-context Huffman and ECB sits at $Q=8$, with ECB outperforming Huffman by up to $27$ percentage points at $Q=32$; on Kodak photographs the crossover shifts below the tested range and ECB wins across the swept $Q$ values, with the gap growing from $0.031$ to $0.113$ bits per symbol.
    \item A mechanism-level finding: HuffmanPos, which keeps Huffman's codewords (and therefore its per-symbol bin count) but allocates one M-coder context per bin position, beats ECB on $12$ of $15$ source cells and loses by at most $0.56$pp on the other three. This is direct evidence that the dominant rate-reducing mechanism at low entropy is per-bin-position context adaptation, not the smaller bin budget of ECB.
    \item A description (without implementation) of an interleaved single-pass ECB decoder that would close the $O(N \cdot m)$ decode-cost gap to UEG.
\end{itemize}

The remainder of the paper is organized as follows. Section~\ref{sec:background} reviews CABAC, the four binarization configurations, and the entropy-conservation property. Section~\ref{sec:related} surveys related work. Section~\ref{sec:impl} describes the implementation. Section~\ref{sec:method} details the benchmark methodology. Section~\ref{sec:results} presents results. Section~\ref{sec:discussion} discusses the crossover, the context-allocation mechanism, and the rate-versus-decoder-complexity trade-off. Section~\ref{sec:limits} states limitations. Section~\ref{sec:conclusion} concludes.

\section{Background}
\label{sec:background}

\subsection{CABAC and the M-coder}

CABAC~\cite{marpe2003cabac} compresses sequences of binary decisions (bins) through a binary arithmetic coder, the M-coder, using adaptive probability models. Arithmetic coding itself dates to Witten, Neal, and Cleary~\cite{witten1987arithmetic}; the M-coder is its multiplication-free, table-driven specialization standardized in H.264/AVC. Each bin is routed through one of three modes: \emph{regular} mode, where the bin is coded against a context model that tracks the most probable symbol (MPS) and a state index parameterizing the probability of the least probable symbol (LPS); \emph{bypass} mode, where the bin is coded as a uniform 50/50 event without context tracking; and \emph{terminate} mode, used to signal end-of-stream and to flush the encoder.

A context model is a tuple $(\sigma, \mathrm{MPS})$ where $\sigma \in [0, 63]$ is the state index and $\mathrm{MPS} \in \{0, 1\}$ is the current most probable symbol. On each coded bin, the M-coder consults a probability table $\mathrm{RANGE\_TAB\_LPS}$ (ITU-T Table 9-44) indexed by $(\sigma, (\mathrm{codIRange} \gg 6) \,\&\, 3)$ to obtain the subrange allocated to the LPS, updates the coding interval, and transitions $\sigma$ via the state-transition tables (Table 9-45). The context adapts to local symbol statistics over time without requiring an explicit probability estimate.

\subsection{Binarization Schemes}

\textbf{UEG.} The H.264 standard binarization for residual magnitudes is the concatenation of a Truncated Unary prefix and a $k$-th order Exp-Golomb suffix. Magnitudes below the cutoff $u_{\text{Coff}}$ are coded with TU alone; magnitudes at or above the cutoff use TU of length $u_{\text{Coff}}$ followed by EG-$k$ on the remainder. The prefix is routed through regular mode (one context per position); the suffix uses bypass. For signed sources, a sign bit is appended in bypass mode.

\textbf{Canonical Huffman.} For an alphabet with empirical frequencies $\{f_i\}$, Huffman's algorithm~\cite{huffman1952method} produces a prefix code minimizing expected codeword length subject to integer-length constraints. The canonical form assigns codewords by length in a deterministic lexicographic order, allowing the codebook to be stored compactly. We evaluate two CABAC variants of Huffman: one with a single M-coder context shared across all bin positions (Huffman), and one with one context per bin position (HuffmanPos). Both produce identical codewords; only the context allocation differs.

\textbf{Entropy-Conserving Binarization (ECB).} Srivastava~\cite{srivastava2014} introduced a binarization scheme that maps an $m$-ary source $X$ over alphabet $\{x_1, \ldots, x_m\}$ into $m-1$ binary strings $Y_1, \ldots, Y_{m-1}$. The construction is sequential: $Y_1$ marks the positions of $x_1$ in $X$; the remaining positions form a reduced sequence over $\{x_2, \ldots, x_m\}$, against which $Y_2$ marks the positions of $x_2$; and so on. The final binary string $Y_{m-1}$ is fully determined.

The central property is the \emph{entropy-conservation theorem}:

\begin{equation}
N \cdot H(Y) = \sum_{i=1}^{m-1} N_i \cdot H(X_i)
\label{eq:entropy_conservation}
\end{equation}

where $N$ is the source length, $H(Y)$ is the joint entropy of the $m-1$ binary strings normalized to per-source-symbol bits, $N_i$ is the length of the $i$-th binary string, and $H(X_i)$ is the entropy of $X_i$. The construction runs in $O(N)$ time and space~\cite{srivastava2014}.

\subsection{Notation}

Throughout, $H(X)$ denotes the empirical first-order Shannon entropy of $X$, $H(X) = -\sum_i p_i \log_2 p_i$, where $p_i$ are observed symbol frequencies. We use this as the reference rate for measuring coding overhead in Section~\ref{sec:method}.

\section{Related Work}
\label{sec:related}

CABAC's design rationale~\cite{marpe2003cabac} balanced rate efficiency against decoder throughput. Sze and Budagavi~\cite{sze2012hevc} analyzed CABAC throughput bottlenecks in HEVC and motivated several binarization simplifications adopted in the HEVC standard, including reduced Rice-parameter Exp-Golomb suffixes and a smaller fraction of regular-mode bins relative to H.264. VVC~\cite{bross2021vvc} retained CABAC and extended its bin-level parallelism without fundamentally revisiting the binarization design space. The H.264 UEG binarization we use as a baseline is therefore representative of the design point that has shaped two subsequent standards.

Within H.264/AVC itself, the question of which binarization is optimal for a given source class has received limited empirical attention since the original standard work. Srivastava's entropy-conserving construction~\cite{srivastava2014} is to our knowledge the only published alternative with a provable optimality property, and its integration into a full arithmetic-coding pipeline (rather than a standalone source-coding evaluation) has not been previously reported.

A parallel line of work in learned entropy coding~\cite{mentzer2018conditional,mentzer2019l3c,balle2018variational} replaces hand-designed binarizations and context models with neural networks that estimate per-symbol probability distributions conditioned on previously coded data. These systems achieve state-of-the-art rate on natural-image distributions but at substantially higher encode and decode cost than CABAC; their probability estimates are typically passed to an arithmetic coder operating directly on the multi-symbol alphabet, sidestepping binarization entirely. The present work is complementary: it evaluates classical binarization choices within a fixed, well-understood arithmetic-coding backend, isolating the contribution of the binarization-and-context-allocation layer that learned approaches have abstracted away.

\section{Implementation}
\label{sec:impl}

We implement CABAC in Java following ITU-T H.264 Section 9.3~\cite{itu2003h264}. The implementation comprises the M-coder, four pluggable binarizers, and a benchmark harness. Source code, build configuration, and raw benchmark output are open\footnote{Code repository: \url{https://github.com/vinamras19/cabac-binarization-study}.}.

\subsection{M-coder}

The arithmetic encoder and decoder maintain the coding interval as a $(\mathrm{codIRange}, \mathrm{codILow})$ pair on the encoder and $(\mathrm{codIRange}, \mathrm{codIOffset})$ on the decoder, initialized to $510$ per the standard. The encoder reads probability table $\mathrm{RANGE\_TAB\_LPS}$ (Table 9-44) at index $(\sigma, (\mathrm{codIRange} \gg 6) \,\&\, 3)$ to obtain the LPS subrange, updates the context state via $\mathrm{TRANS\_IDX\_MPS}$ and $\mathrm{TRANS\_IDX\_LPS}$ (Table 9-45), and renormalizes when $\mathrm{codIRange} < 256$. Outstanding bits are deferred through a $\mathrm{bitsOutstanding}$ counter and emitted on the next determined bit. Three coding modes are exposed: regular, bypass, and terminate. The decoder mirrors the encoder structure, reading $9$ bits to initialize $\mathrm{codIOffset}$ and renormalizing in lockstep.

\subsection{Pluggable Binarizers}

All four binarizers expose a uniform interface providing \texttt{encode} and \texttt{decode} methods over an arithmetic-coding backend. This guarantees that the schemes are evaluated through the same M-coder, with no implementation differences in the arithmetic-coding layer.

\textbf{UEG.} Uses $u_{\text{Coff}} = 14$ context models for the Truncated Unary prefix and bypass mode for the Exp-Golomb suffix and sign bit. EG-$k$ order is fixed at $k=0$.

\textbf{ECB.} The encoder iterates over $m-1$ passes, on each pass emitting bins through one context model (allocated per binary string). The decoder maintains a $\mathrm{resolved}$ flag array and iterates over positions $m-1$ times in the worst case; this implementation is correct but yields $O(N \cdot m)$ decode complexity. An interleaved single-pass variant would close this gap; we describe its structure in Section~\ref{sec:limits}.

\textbf{Huffman.} The encoder builds a Huffman tree from observed frequencies, assigns canonical codeword lengths, and emits all codeword bits through a single shared M-coder context.

\textbf{HuffmanPos.} Identical codeword construction to Huffman, but the encoder allocates one M-coder context per bin position (sized to the maximum codeword length) and routes the $i$-th bin of each codeword through context $i$. The decoder mirrors this by tracking a position counter that resets on each completed codeword match. This isolates the effect of per-position context adaptation from the choice of binarization, since the bin count emitted per source symbol is identical between Huffman and HuffmanPos.

\section{Methodology}
\label{sec:method}

We evaluate the four binarization configurations through the shared M-coder of Section~\ref{sec:impl} on three source classes: synthetic distributions, quantized DCT residuals from a procedural test image, and quantized DCT residuals from a set of natural photographs. Every trial is encoded, decoded, and verified bit-exact against the source.

\subsection{Sources}
\label{sec:sources}

\textit{Synthetic distributions.} Five sources at $N=10{,}000$ samples per trial: uniform over $[0,8)$; biased over $[0,8)$ with $P(0)=0.7$; truncated geometric over $[0,8)$ with parameter $p=0.6$; signed Laplacian with scale $b=2$ (narrow); and signed Laplacian with $b=8$ (wide). The geometric source is truncated to alphabet size rather than unbounded. Laplacian samples are integer-rounded.

\textit{Procedural DCT residuals.} A procedurally generated $256 \times 256$ test image consisting of a sinusoidal gradient, six Gaussian bumps, four sharp rectangles, and additive Gaussian sensor noise. Each $8 \times 8$ block is centered by subtracting $128$, transformed by DCT-II, quantized by step $Q$, and zig-zag-scanned, yielding $65{,}536$ integer coefficients per trial. We sweep $Q \in \{2, 4, 8, 16, 32\}$, producing alphabets ranging from $m=571$ at $Q=2$ to $m=55$ at $Q=32$.

\textit{Kodak natural photographs.} All 24 images of the Kodak True Color Image Suite~\cite{kodak_suite} (kodim01 through kodim24). RGB images are converted to BT.601 luma ($Y' = 0.299R + 0.587G + 0.114B$) and processed through the same DCT pipeline as the procedural image, yielding approximately $393{,}216$ coefficients per image. The same $Q$ values are swept; for each $(Q, \text{scheme})$ cell, results are aggregated across the 24 images, and standard deviation is reported across images rather than seeds.

\subsection{Benchmark Protocol}
\label{sec:protocol}

Synthetic and procedural cells run $50$ seeds in sequential order; Kodak cells run one trial per image. Before measurement, we warm the JVM with nine Laplacian trials at $N=1{,}000$. Encode and decode latencies are captured with \texttt{System.nanoTime()} around each call; we report mean and standard deviation per cell. Binarizers receive the parameters natural to their construction: UEG uses $u_{\text{Coff}}=14$ and $k=0$ throughout, with signed mode auto-detected per source; ECB uses empirical-frequency-descending symbol order; both Huffman variants build codebooks per-trial from observed frequencies, and codebook serialization cost is not counted in the reported bit rate.

The four schemes use the context-allocation strategies described in Section~\ref{sec:impl}. UEG allocates $14$ contexts for its TU prefix and uses bypass mode for the suffix; ECB allocates $m-1$ contexts (one per binary string); Huffman uses one shared context across all positions; HuffmanPos allocates one context per codeword bit position (up to the maximum codeword length).

The full benchmark suite is reproducible from a single command (\texttt{mvn exec:java -Dexec.mainClass=com.cabac.BenchmarkSuite}) and runs against Java~$17$. Synthetic and procedural seeds are integers in sequential order; the Kodak image set is fixed at the 24 files comprising the full suite. The harness writes the per-cell aggregates to \texttt{benchmark/benchmark-results.csv}, from which all tables and figures in this paper are produced.

\subsection{Metric and Verification}
\label{sec:metric}

We report \emph{coding overhead} as

\begin{equation}
\eta(X) = \frac{R(X) - H(X)}{H(X)} \times 100\%
\label{eq:eta}
\end{equation}

where $R(X)$ is the encoded bit rate in bits per source symbol and $H(X) = -\sum_i p_i \log_2 p_i$ is the empirical first-order Shannon entropy. Lower is better; $\eta = 0$ corresponds to an ideal code at the first-order entropy. Note that $\eta$ can be negative when the adaptive M-coder captures conditional structure (within-block or block-to-block correlations in DCT residuals) that the plug-in marginal estimator does not. We return to this in Section~\ref{sec:discussion}.

Every trial is round-tripped: the encoded bit stream is decoded back into a symbol array and compared bit-exact to the source. A cell passes only if every trial round-trips successfully.

\section{Results}
\label{sec:results}

\subsection{Synthetic Distributions}

Table~\ref{tab:synthetic} reports coding overhead $\eta$ across the five synthetic sources. No scheme dominates: each binarization wins on the source class to which it is structurally aligned, and HuffmanPos lands within $0.5$ percentage points of the leader on every source. Huffman is the best base scheme on the uniform alphabet (where 3-bit codewords match $\log_2(8)$ exactly); HuffmanPos extends the lead by $0.2$pp, an effect smaller than the $0.02$ bits/symbol seed-to-seed standard deviation. UEG wins on the narrow Laplacian, for which TU+EG-$k$ was designed. ECB is the best base scheme on the biased and geometric sources, where the most-frequent-first symbol order maps the dominant value to a binary string the M-coder adapts to cheaply. On the wide Laplacian, HuffmanPos overtakes single-context Huffman by $0.23$pp. Standard deviation across $50$ seeds is at most $0.02$ bits/symbol for all cells; the reported differences exceed sampling noise on the biased, geometric, and Laplacian sources.

\begin{table}[htbp]
\caption{Coding overhead $\eta$ on synthetic sources, $N=10{,}000$, $50$ seeds. Best per row in bold.}
\label{tab:synthetic}
\centering
\begin{tabular}{lcccc}
\toprule
Source & UEG & ECB & Huffman & HuffmanPos \\
\midrule
Uniform $[0,8)$ & $3.38\%$ & $3.16\%$ & $1.86\%$ & $\mathbf{1.68\%}$ \\
Biased $[0,8)$, $P(0)=0.7$ & $3.01\%$ & $2.78\%$ & $6.78\%$ & $\mathbf{2.59\%}$\\
Geometric $[0,8)$, $p=0.6$ & $2.22\%$ & $\mathbf{2.15\%}$ & $2.20\%$ & $2.17\%$ \\
Laplacian, $b=2$ & $\mathbf{1.91\%}$ & $3.03\%$ & $2.85\%$ & $2.39\%$ \\
Laplacian, $b=8$ & $6.21\%$ & $6.72\%$ & $2.34\%$ & $\mathbf{2.11\%}$ \\
\bottomrule
\end{tabular}
\end{table}

Single-context Huffman's $6.78\%$ overhead on the biased source, despite its symbol-level optimality, contrasts sharply with HuffmanPos's $2.59\%$ on the same source. The two schemes share identical codewords, so the $4.19$pp gap is attributable entirely to context allocation: replacing one shared M-coder context with one context per bin position recovers most of the loss. This previews the central mechanism we examine in the DCT sweeps.

\subsection{DCT Quantization Sweep: Procedural Image}

Table~\ref{tab:dct} reports $\eta$ on quantized DCT residuals from the procedural test image, swept over $Q \in \{2, 4, 8, 16, 32\}$. As $Q$ increases, the source shifts from rich (large alphabet, near-stationary AC distribution) to sparse (small alphabet, zero-dominated).

\begin{table}[htbp]
\caption{Coding overhead $\eta$ on procedural DCT residuals by quantization step $Q$. $|\mathcal{X}|$ is alphabet size. $65{,}536$ coefficients per trial, $50$ seeds. Best per row in bold.}
\label{tab:dct}
\centering
\begin{tabular}{ccccccc}
\toprule
$Q$ & $|\mathcal{X}|$ & $H(X)$ & UEG & ECB & Huffman & HuffmanPos \\
\midrule
$2$ & $571$ & $3.801$ & $3.95\%$ & $6.86\%$ & $2.22\%$ & $\mathbf{2.05\%}$ \\
$4$ & $350$ & $2.830$ & $4.08\%$ & $5.00\%$ & $2.34\%$ & $\mathbf{1.66\%}$ \\
$8$ & $193$ & $1.923$ & $4.04\%$ & $3.54\%$ & $4.54\%$ & $\mathbf{1.62\%}$ \\
$16$ & $105$ & $1.001$ & $4.87\%$ & $3.31\%$ & $11.92\%$ & $\mathbf{2.01\%}$ \\
$32$ & $55$ & $0.394$ & $3.15\%$ & $-0.69\%$ & $26.21\%$ & $\mathbf{-0.83\%}$ \\
\bottomrule
\end{tabular}
\end{table}

The single-context-Huffman vs.\ ECB crossover sits at $Q=8$. Below it, Huffman dominates by approximately two percentage points. Above it, ECB takes over, reaching $27$ percentage points below Huffman at $Q=32$. UEG occupies a stable $3$--$5\%$ band across the full sweep, never excelling but never failing catastrophically.

HuffmanPos changes the picture meaningfully. Using the same codewords as Huffman, and therefore committing the same number of bins per source symbol, HuffmanPos wins across the entire procedural sweep. At $Q=32$, its rate (\,$0.391$ bps,\,$\eta = -0.83\%$) is statistically indistinguishable from ECB's ($0.391$ bps,\,$\eta = -0.69\%$). The bin-count-based explanation for Huffman's failure, namely that integer codeword lengths force more bins per source symbol than ECB's geometrically shrinking $m-1$ streams, does not apply to HuffmanPos, which has the same bin count as Huffman but per-position contexts. The natural reinterpretation is that the bin-count gap matters less than the context-allocation strategy at low entropy. We develop this argument in Section~\ref{sec:discussion}.

Both ECB and HuffmanPos at $Q=32$ land below the empirical first-order entropy ($\eta = -0.69\%$ and $-0.83\%$ respectively). This is not a violation of source-coding bounds; it reflects two effects, examined in Section~\ref{sec:discussion}.

\subsection{DCT Quantization Sweep: Kodak Natural Photographs}

Table~\ref{tab:kodak} reports $\eta$ on quantized DCT residuals from the 24 Kodak photographs, swept over the same $Q$ values.

\begin{table}[htbp]
\caption{Coding overhead $\eta$ on Kodak DCT residuals by quantization step $Q$. $|\mathcal{X}|$ is the alphabet size aggregated across 24 photographs; $\sim 393{,}216$ coefficients per image. Best per row in bold.}
\label{tab:kodak}
\centering
\begin{tabular}{ccccccc}
\toprule
$Q$ & $|\mathcal{X}|$ & $H(X)$ & UEG & ECB & Huffman & HuffmanPos \\
\midrule
$2$ & $834$ & $4.053$ & $-1.18\%$ & $-1.13\%$ & $-0.38\%$ & $\mathbf{-3.60\%}$ \\
$4$ & $445$ & $3.097$ & $-3.66\%$ & $-4.22\%$ & $-1.80\%$ & $\mathbf{-5.48\%}$ \\
$8$ & $227$ & $2.208$ & $-6.03\%$ & $-7.26\%$ & $-3.64\%$ & $\mathbf{-7.67\%}$ \\
$16$ & $114$ & $1.478$ & $-7.31\%$ & $\mathbf{-8.94\%}$ & $-2.55\%$ & $-8.89\%$ \\
$32$ & $58$ & $0.909$ & $-7.64\%$ & $\mathbf{-9.10\%}$ & $3.37\%$ & $-8.54\%$ \\
\bottomrule
\end{tabular}
\end{table}

\begin{figure*}[t]
\centering
\includegraphics[width=\textwidth]{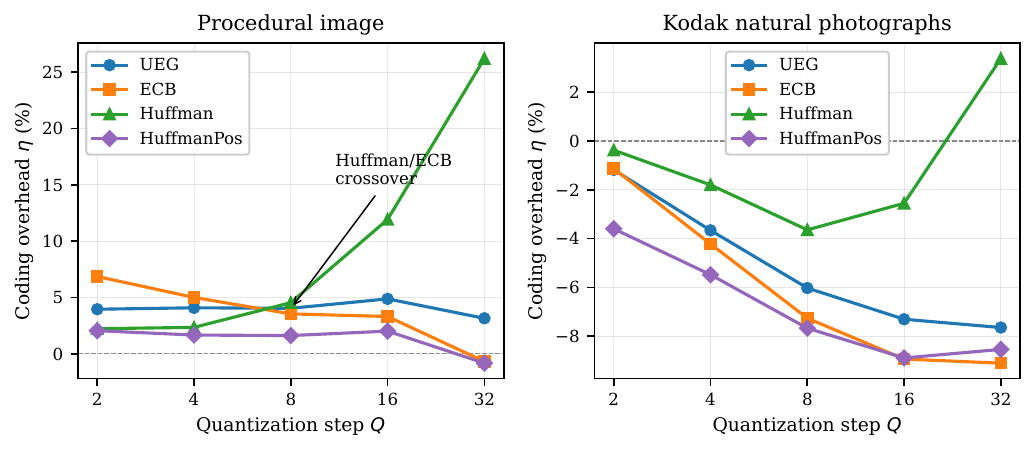}
\caption{Coding overhead $\eta$ vs.\ quantization step $Q$ on the procedural test image (left panel) and the full 24-image Kodak suite (right panel). The Huffman/ECB crossover at $Q=8$ on the procedural image does not generalize to natural images; on Kodak, ECB beats single-context Huffman at every tested $Q$ and absolute overheads shrink to a narrower (and largely negative) range. HuffmanPos tracks the leading scheme across both panels, demonstrating that per-position context allocation closes most of single-context Huffman's gap to ECB.}
\label{fig:eta_vs_q}
\end{figure*}

Two structural differences from the procedural sweep are visible in Fig.~\ref{fig:eta_vs_q}. First, the crossover position shifts: ECB beats single-context Huffman at every tested $Q$, with the gap growing monotonically from $0.031$ bps at $Q=2$ to $0.113$ bps at $Q=32$. The procedural crossover at $Q=8$ does not generalize. The ECB-vs-UEG crossover similarly moves from between $Q=4$ and $Q=8$ on the procedural image to between $Q=2$ and $Q=4$ on Kodak. Second, absolute overheads shrink: all four schemes code below the empirical first-order entropy at $Q=2$ (single-context Huffman by the smallest margin, $0.38\%$), and UEG, ECB, and HuffmanPos stay below it at every tested $Q$; only single-context Huffman returns to positive overhead, reaching $+3.37\%$ at $Q=32$. The negative-overhead pattern observed at procedural $Q=32$ is the rule on natural images, not the exception. We discuss both phenomena in Section~\ref{sec:discussion}.

HuffmanPos and ECB are statistically indistinguishable on Kodak. HuffmanPos leads at $Q \in \{2, 4, 8\}$ by $0.4$--$2.5$ percentage points; ECB leads narrowly at $Q \in \{16, 32\}$ by $0.05$--$0.6$ percentage points. Both schemes dominate single-context Huffman by similar margins, reaching $-8.5$ to $-9.1\%$ at $Q=32$ where single-context Huffman pays $+3.37\%$.

Per-image variance is substantial. At $Q=2$, bits per symbol across the 24 images has standard deviation $0.76$ against a mean of $4.01$. Different photographs have very different statistics. Scheme-to-scheme differences are smaller than image-to-image differences in absolute terms, but the scheme ordering is consistent across images and $Q$ values.

\subsection{Round-trip Correctness}

All $2{,}480$ trials decode bit-exact to their source array: $5$ synthetic sources $+$ $5$ procedural DCT $Q$ values, $\times$ $4$ schemes, $\times$ $50$ seeds, plus $5$ Kodak $Q$ values $\times$ $4$ schemes $\times$ $24$ images. No mismatches were observed across the parameter ranges tested.

\subsection{Decoder Latency}

ECB's rate efficiency at high $Q$ comes at decoder cost. Table~\ref{tab:decode} compares decoder latency across sources at varying alphabet sizes $m$.

\begin{table}[htbp]
\caption{Decoder latency by source and scheme, mean across trials. ECB cost grows approximately linearly in alphabet size $m$ while UEG, Huffman, and HuffmanPos scale visibly more slowly (see Fig.~\ref{fig:decoder_latency}); HuffmanPos pays a roughly constant-factor premium over Huffman (6--17\% across the cells shown); Kodak per-trial times scale with the larger coefficient count per image.}
\label{tab:decode}
\centering
\begin{tabular}{lccccc}
\toprule
Source & $m$ & UEG & ECB & Huffman & HuffmanPos \\
\midrule
Laplacian $b=8$ & $111$ & $0.77$ & $2.03$ & $1.92$ & $2.16$ \\
Procedural $Q=2$ & $571$ & $3.21$ & $24.45$ & $8.58$ & $9.74$ \\
Procedural $Q=8$ & $193$ & $1.72$ & $9.55$ & $4.66$ & $5.42$ \\
Procedural $Q=32$ & $55$ & $0.57$ & $2.66$ & $2.48$ & $2.63$ \\
Kodak $Q=2$ & $834$ & $20.54$ & $206.68$ & $56.51$ & $63.39$ \\
Kodak $Q=8$ & $227$ & $10.55$ & $56.92$ & $29.32$ & $33.53$ \\
Kodak $Q=32$ & $58$ & $5.04$ & $17.04$ & $17.16$ & $18.87$ \\
\bottomrule
\end{tabular}
\\[2pt]
\footnotesize All values in milliseconds per trial.
\end{table}

ECB decode ranges from $4.7\times$ ($Q=32$) to $7.6\times$ ($Q=2$) slower than UEG on procedural DCT sources and reaches $10.1\times$ on Kodak at $Q=2$ ($m=834$). The ratio grows approximately linearly with $m$. On smaller-alphabet sources such as the wide Laplacian ($m=111$), the ratio is closer to $2.6\times$. Kodak absolute times are approximately $6\times$ larger than the procedural equivalent because each trial processes about $6\times$ more coefficients; the per-scheme ratios are broadly preserved but amplified at large $m$, attributable to cache pressure across the $m-1$ array passes. HuffmanPos decode is $6$--$28\%$ slower than single-context Huffman depending on the source ($6$--$17\%$ across the DCT cells of Table~\ref{tab:decode}; the small-alphabet synthetic sources reach the high end), attributable to maintaining a position-indexed context array and resetting the position counter on each completed codeword; the per-bin arithmetic decoding cost itself is identical. The ECB cost is attributable to our $O(N \cdot m)$ decoder implementation, which iterates $m-1$ times through the symbol array; we describe an interleaved single-pass variant that would close this gap in Section~\ref{sec:limits}.

\section{Discussion}
\label{sec:discussion}

\subsection{The Sparsity-Driven Crossover and Its Image Dependence}

The procedural $Q$-sweep revealed a deterministic transition between two regimes at $Q=8$. Below the crossover, the residual distribution was rich (large alphabet, high entropy, near-stationary AC distribution) and single-context Huffman codes exploited their symbol-level optimality, achieving $\eta \approx 2.2$--$2.3\%$. Above the crossover, more DCT coefficients quantized to zero, the alphabet collapsed toward zeros and $\pm 1$, and single-context Huffman's overhead ballooned to $26.21\%$ at $Q=32$ while ECB dropped to $\eta = -0.69\%$ and HuffmanPos to $\eta = -0.83\%$.

The Kodak sweep shows that the position of this crossover is not universal. On natural photographs the crossover sits below the tested range, and single-context Huffman never wins. The mechanism is the same in both cases (Section~\ref{sec:mechanism}); what shifts is the threshold $Q$ at which the mechanism takes hold. Real images have lower $H(X)$ at every $Q$ than the procedural image because of stronger low-frequency concentration: a sinusoidal gradient with Gaussian bumps does not concentrate energy in low spatial frequencies the way a natural photograph does, so the procedural residual distribution remains rich (high entropy) at lower $Q$ values. The threshold at which a single shared context becomes inadequate to compress Huffman's bin stream is therefore reached at different $Q$ values on different image classes.

\subsection{Mechanism: Bin Count vs.\ Context Allocation}
\label{sec:mechanism}

Two factors determine the rate of a binarization-plus-context-allocation choice driven through CABAC: the number of bins emitted per source symbol, and how those bins are partitioned into M-coder contexts. The four schemes under study span the relevant corners:

\begin{itemize}
    \item \textbf{Bin count and context allocation differ:} UEG vs.\ ECB vs.\ Huffman, where each binarization has a different per-symbol bin count, and each uses a different context-allocation strategy.
    \item \textbf{Bin count identical, context allocation differs:} Huffman vs.\ HuffmanPos. Same codewords, hence same number of bins per source symbol. Different contexts.
\end{itemize}

The Huffman vs.\ HuffmanPos comparison therefore isolates the effect of context allocation. At procedural $Q=32$, the two schemes commit identical bin counts (Huffman's codebook produces the same expected codeword length under either configuration). Yet the encoded rates differ by $21\%$: $0.497$ bps for single-context Huffman vs.\ $0.391$ bps for HuffmanPos. The full $21\%$ difference is attributable to context allocation; bin count plays no role.

\begin{figure}[t]
\centering
\includegraphics[width=\columnwidth]{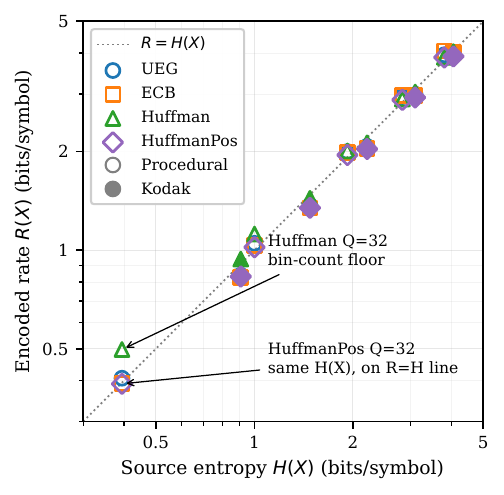}
\caption{Empirical encoded rate $R(X)$ vs.\ source entropy $H(X)$ across all DCT cells (procedural and Kodak). The dotted reference line marks $R = H(X)$. All four schemes track the reference closely at high entropy. At low entropy (procedural $Q=32$, leftmost cluster), single-context Huffman shows visible positive departure (the bin-count floor argument applies to it), while HuffmanPos, with the same codewords but per-position contexts, sits on the line at the same $H(X)$. Both annotated points share alphabet size and codebook; only the context allocation differs.}
\label{fig:rate_vs_entropy}
\end{figure}

Fig.~\ref{fig:rate_vs_entropy} shows this directly. Across all DCT cells (procedural and Kodak), encoded rate $R(X)$ is plotted against source entropy $H(X)$. The dotted reference line marks $R = H(X)$. At procedural $Q=32$, single-context Huffman sits visibly above the line at the lowest-$H(X)$ region of the figure; HuffmanPos sits on the line at the same $H(X)$, despite emitting the same bins. The figure is the cleanest empirical statement of the context-allocation finding: same source, same bin count, two different points.

ECB's structural advantage over single-context Huffman is real but operates through a different lever. Recall the per-symbol bin counts:
\begin{equation}
\bar{\ell}_{\text{Huff}} = \sum_{i=1}^{m} p_i \, \ell_i, \qquad
\bar{\ell}_{\text{ECB}} = \sum_{i=1}^{m-1} \Pr\!\left[X \notin \{x_1, \ldots, x_{i-1}\}\right],
\label{eq:bin_count}
\end{equation}
where $\ell_i$ is the Huffman codeword length for symbol $x_i$. As $p_{\max} \to 1$, $\bar{\ell}_{\text{ECB}} \to 1$ from above while $\bar{\ell}_{\text{Huff}}$ is bounded below by $\ell_1 \geq 1$ with the $m-1$ non-dominant codewords contributing multi-bit terms. ECB's smaller bin budget, combined with one context per binary string, gives it a structural rate advantage at low entropy. HuffmanPos shows that this advantage can be replicated with Huffman's codewords if the contexts are allocated similarly. The two schemes reach the same rate at procedural $Q=32$ through different routes: ECB by emitting fewer bins each routed through its own context; HuffmanPos by emitting the same bins as single-context Huffman but routing each bin position through its own context.

The unified statement: at low source entropy, the M-coder can only adapt as quickly as the per-position bin statistics allow. A single context blends bin positions with very different local statistics together, slowing adaptation and inflating rate. Per-position contexts (whether across binary streams in ECB, or across codeword positions in HuffmanPos) let the M-coder adapt at the natural granularity of the bin stream's structure. Bin count contributes to absolute rate; context allocation determines whether the M-coder can compress that bin count near its entropy.

\subsection{Below the First-Order Entropy}

At procedural $Q=32$, both ECB and HuffmanPos code \emph{below} $H(X)$ ($\eta = -0.69\%$ and $-0.83\%$ respectively). On the Kodak sweep, UEG, ECB, and HuffmanPos code below $H(X)$ at every tested $Q$. This warrants explanation, as it appears at first to violate source-coding bounds.

Two effects account for the gap. First, the plug-in Shannon entropy estimator $\hat{H}(X) = -\sum_i \hat{p}_i \log_2 \hat{p}_i$ is known to be downward-biased when the alphabet is large relative to the sample count~\cite{paninski2003estimation}; the leading-order correction, due to Miller~\cite{miller1955note}, is $(m-1)/(2N \ln 2)$. At procedural $Q=32$ ($m=55$, $N=65{,}536$), this bias correction evaluates to approximately $6 \times 10^{-4}$ bits/symbol and accounts for roughly $20\%$ of the observed gap. On Kodak, with $N \approx 393{,}216$ per image, Miller-Madow bias is negligible.

Second, and more substantively, $H(X)$ as we compute it is the marginal first-order entropy, which assumes the source is i.i.d. DCT residuals are not i.i.d.: within-block coefficients have correlation along the zig-zag scan, and block-to-block coefficients are spatially correlated through the underlying image structure. The M-coder's adaptive contexts, evolving over the source, capture some of this conditional structure. The schemes are therefore coding closer to the \emph{conditional} entropy of the source than to its marginal entropy. The marginal estimator undershoots the achievable rate by a margin that grows with how much conditional structure the source carries; Kodak photographs ($1/f$ power spectrum, smooth correlated regions, structured edges) carry substantially more conditional structure than the procedural image, so the sub-entropy pattern is amplified there. None of this violates source-coding bounds~\cite{cover2006elements}; it merely exposes the limits of memoryless entropy estimation.

\subsection{UEG's Stable Middle Band}

UEG occupies a $3$--$5\%$ overhead band on the procedural sweep, and a $-1$ to $-8\%$ band on Kodak, neither best nor worst at any tested $Q$. This stability is the property that made UEG the H.264 default~\cite{marpe2003cabac}, and that HEVC and VVC preserved with refinements~\cite{sze2012hevc,bross2021vvc}. The TU prefix with $u_{\text{Coff}} = 14$ contexts provides per-position adaptation; the EG-$k$ suffix in bypass mode handles large-magnitude tails without context overhead. UEG is the most predictable of the four schemes across operating points and image classes, even though it is dominated on either side of the crossover by a more specialized scheme.

\begin{figure*}[!t]
\centering
\includegraphics[width=\textwidth]{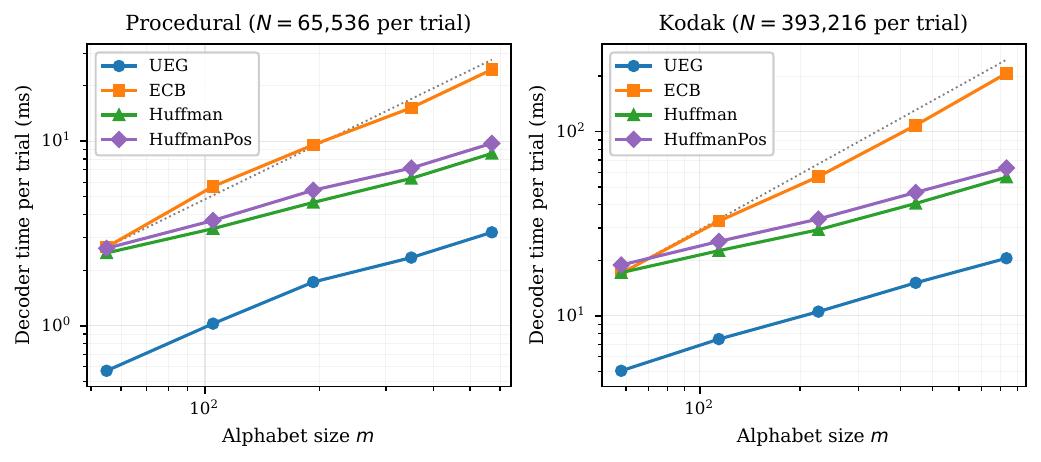}
\caption{Per-trial decoder latency vs.\ alphabet size $m$ on log-log axes, for procedural (left, $N{=}65{,}536$ coefficients per trial) and Kodak (right, $N{=}393{,}216$ coefficients per trial). The dotted reference has slope $1$ in $m$ and is anchored at ECB's leftmost point. ECB's curve tracks the reference closely in both panels (empirical $\propto m^{0.92}$, slightly sub-linear because later decoder passes operate on fewer remaining unresolved positions); UEG, Huffman, and HuffmanPos scale visibly more slowly. HuffmanPos tracks $6$--$17\%$ above Huffman across the cells shown, a constant-factor cost. Absolute times differ between panels because of the different sample count per trial, but the relative scaling is preserved.}
\label{fig:decoder_latency}
\end{figure*}

\subsection{Rate versus Decoder Complexity}

ECB's rate advantage at high $Q$ is real, but our current implementation pays for it on the decode side. The $O(N \cdot m)$ decoder, which makes $m-1$ passes over the symbol array, runs $7.6\times$ slower than UEG at procedural $Q=2$ ($m=571$). On Kodak the ratio amplifies further: ECB decode at $Q=2$ ($m=834$) runs $10.1\times$ slower than UEG. Fig.~\ref{fig:decoder_latency} plots decoder latency against alphabet size $m$ on log-log axes. ECB's curve tracks closely to the dotted slope-$1$ reference across both procedural and Kodak source classes, consistent with approximately linear scaling in $m$; later passes operate on fewer remaining unresolved positions, which produces the slight sub-linearity. UEG, Huffman, and HuffmanPos scale visibly more slowly. HuffmanPos tracks $6$--$17\%$ above Huffman across the $m$ range shown, a constant-factor cost from maintaining a position-indexed context array, not a scaling penalty. The near-linear ECB cost is an artifact of the multi-pass decoder construction, not an algorithmic property of ECB. An interleaved single-pass decoder that consumes one bin from each of the $m-1$ streams per source position would achieve $O(N)$ complexity at the cost of more complex state management. We sketch this construction in Section~\ref{sec:limits}.

\section{Limitations}
\label{sec:limits}

We disclose the following limitations.

\textbf{Limited image diversity.} The DCT residual experiments use a procedurally generated $256 \times 256$ image and the 24 Kodak natural photographs. Other content classes (high-resolution modern photos, screen content, video frames) are not tested. The mechanism-level explanation in Section~\ref{sec:mechanism} suggests the scheme ordering generalizes, but absolute overheads will shift further on different image classes.

\textbf{ECB decoder is $O(N \cdot m)$, but a single-pass variant is feasible.} The current encoder emits the $m-1$ binary streams as concatenated blocks ($Y_1$ in full, then $Y_2$, and so on), which forces the decoder to make $m-1$ passes over the symbol array. A single-pass implementation requires the encoder to be restructured to interleave the streams: at each unresolved source position $j$ (in encoder-side natural order), the encoder emits one bin from the currently active stream; if that bin is $1$ the position is resolved and skipped on subsequent visits, otherwise the position is revisited under the next stream. The decoder mirrors this exactly: at each unresolved source position $j$, consume one bin from the active stream (initially $Y_1$); if the bin is $1$, mark $X_j = x_i$ where $i$ is the active stream index; otherwise advance position $j$ to the next stream. The same $(context, bin)$ pairs are encoded in a different order, so the total coded length is unchanged. The state per source position is a single integer (the current stream index for that position); total work is $\sum_j (\text{streams visited at position } j) = \sum_i N_i$, which equals the total bin count and is $O(N)$. This is structurally analogous to bin-level interleaving in modern entropy coders and would close the latency gap to UEG; we have not implemented or evaluated this variant in the present work.

\textbf{Huffman codebook serialization not counted.} A deployed Huffman or HuffmanPos codec must store its codebook alongside the compressed stream. Our benchmark omits this cost. The reported overheads on uniform and wide-Laplacian sources are therefore lower bounds. At $N \geq 10{,}000$ the cost is small relative to total rate, but a strict apples-to-apples comparison should include it.

\textbf{Plug-in entropy estimator.} The reference $H(X)$ used to compute $\eta$ is the plug-in Shannon entropy of observed marginal frequencies. This estimator carries Miller-Madow bias at low entropy and large alphabet, and structurally ignores spatial correlation that the M-coder partially captures. The negative-overhead cells throughout the Kodak table and at procedural $Q=32$ should be read as statements that the schemes code near the conditional entropy of the source, not as violations of source-coding bounds.

\textbf{UEG parameters not tuned.} The benchmark uses default $u_{\text{Coff}} = 14$ and $k = 0$ throughout. UEG's gap to Huffman on rich DCT sources may close with parameter search; we did not perform this search.

\textbf{JVM timing.} Encode and decode latencies are measured with \texttt{System.nanoTime()} after JIT warmup, but without fork isolation, GC management, or JMH-grade harnessing. Encode-time standard deviation is small relative to mean for most cells. ECB decode at large $m$ shows higher timing variance attributable to cache behavior across the $m-1$ array passes.

\section{Conclusion}
\label{sec:conclusion}

We integrated the entropy-conserving binarization scheme of Srivastava~\cite{srivastava2014} into a from-scratch H.264/AVC CABAC implementation alongside UEG, canonical Huffman with a single shared context, and a Huffman variant with one context per bin position (HuffmanPos), all sharing a single M-coder backend. Across $2{,}480$ bit-exact-verified trials on synthetic distributions, a procedural DCT image, and the 24-image Kodak True Color Image Suite, no scheme dominates universally. On the procedural image we identified a sparsity-driven crossover at $Q=8$ where ECB overtakes single-context Huffman, reaching $27$ percentage points below at $Q=32$. On Kodak natural photographs the crossover shifts below the tested range: ECB beats single-context Huffman at every $Q$ tested, by margins growing from $0.031$ to $0.113$ bits per symbol.

HuffmanPos isolates the mechanism. It uses identical codewords to single-context Huffman, hence the same per-symbol bin count, but allocates one M-coder context per bin position. HuffmanPos beats ECB on $12$ of the $15$ source cells and loses by at most $0.56$pp on the other three, despite committing the same number of bins as single-context Huffman. This is direct evidence that the rate gap is driven primarily by context allocation strategy over the bin stream, not by the binarization's per-symbol bin count. The bin-count argument that motivates ECB's structural advantage is real, but the dominant mechanism observable in our data is context allocation. Both levers reduce rate at low entropy; the empirical question of which dominates is settled in favor of context allocation by the Huffman-vs-HuffmanPos comparison.

ECB's rate advantage costs $7$ to $10\times$ in decoder latency on large alphabets, traced to an $O(N \cdot m)$ decoder. We sketched an interleaved single-pass variant that would close this gap. HuffmanPos pays a small constant-factor decode premium over Huffman (6--28\% depending on the source), making it the most favorable rate-versus-decoder-cost trade-off among the four schemes on natural images. The natural next steps are (i) implementing the interleaved single-pass ECB decoder to validate the $O(N)$ analysis empirically, (ii) evaluation on broader content classes including high-resolution photographs, screen content, and video frames, and (iii) extending the context-allocation analysis to HEVC and VVC binarization structures~\cite{sze2012hevc,bross2021vvc}. Code and benchmarks are publicly available.

\bibliographystyle{IEEEtran}
\bibliography{references}

\end{document}